\documentclass[twocolumn]{aastex62}

\usepackage{ulem}
\hypersetup{linkcolor=red,citecolor=blue,filecolor=cyan,urlcolor=magenta}
\usepackage{amsmath}
\usepackage{tensor}     
\usepackage{graphicx}   
\usepackage{float}
\usepackage{bm}         
\usepackage{xcolor}     
\usepackage{color}      
\usepackage[section]{placeins} 
\usepackage[utf8]{inputenc} 
\usepackage[flushleft]{threeparttable}
\newcommand{\nc}{\newcommand*} 
\nc{\figurewidth}{3.2in}
\nc{\xbar}{\bar{x}}
\nc{\rhoeq}{\rho_{\mathrm{eq}}}
\nc{\zeq}{z_{\mathrm{eq}}}
\nc{\tla}{\tilde{\lambda}}
\nc{\dt}{\delta}
\nc{\Dt}{\Delta}
\nc{\vj}{\vec{j}}
\nc{\vl}{\vec{l}}
\nc{\hx}{\hat{x}}
\nc{\hy}{\hat{y}}
\nc{\bj}{\bm{j}}
\nc{\mJ}{\mathcal{J}}
\nc{\mP}{\mathcal{P}}
\nc{\Msun}{M_\odot}
\nc{\app}{\approx}
\nc{\av}[1]{\langle #1 \rangle}
\nc{\eq}[1]{Eq.~\eqref{#1}}
\nc{\al}{\alpha}
\nc{\Xstar}{X_{\ast}}
\nc{\seq}{\sigma_{\mathrm{eq}}}
\nc{\fpbh}{f_{\mathrm{pbh}}}
\nc{\vth}{\vec{\theta}}
\nc{\vla}{\vec{\lambda}}
\nc{\vd}{\vec{d}}
\nc{\Mmin}{M_{\mathrm{min}}}
\nc{\rmd}{\mathrm{d}}
\nc{\mmin}{{m_{\mathrm{min}}}}
\nc{\mmax}{{m_{\mathrm{max}}}}
\nc{\mR}{\mathcal{R}}
\nc{\tmR}{\tilde{\mathcal{R}}}
\nc{\s}{\sigma}
\nc{\ogw}{\Omega_{\mathrm{GW}}}
\nc{\addref}{[\textcolor{red}{add ref}] }
\nc{\Om}{\Omega}
\nc{\gpcyr}{\mathrm{Gpc}^{-3}\,\mathrm{yr}^{-1}}
\nc{\Eq}[1]{Eq.~\eqref{#1}}
\nc{\Fig}[1]{Figure~\ref{#1}}
\nc{\Table}[1]{Table~\ref{#1}}
\nc{\lvc}{LIGO/Virgo} 
\nc{\Sec}[1]{Sec.~\ref{#1}}
\nc{\eg}{\textit{e.g.~}}
\nc{\SNR}{\mathrm{SNR}}

\def\({\left(}
\def\){\right)}
\def\[{\left[}
\def\]{\right]}

\def\e{\begin{equation}}
\def\q{\end{equation}}
\def\m{\begin{eqnarray}}
\def\n{\end{eqnarray}}

\received{}
\revised{}
\accepted{}
\published{}
\submitjournal{ApJ}
\begin{document}

\title{Orbit Tomography of  Binary Supermassive Black Holes with Very Long Baseline Interferometry}
	
\author{Yun Fang}
\email{fang.yun@pku.edu.cn}
\affiliation{Kavli Institute for Astronomy and Astrophysics, Peking University, Beijing 100871, China}

\author{Huan Yang}
\email{hyang@perimeterinstitute.ca}
\affiliation{Perimeter Institute for Theoretical Physics, Ontario, N2L 2Y5, Canada}
\affiliation{University of Guelph, Guelph, Ontario N1G 2W1, Canada}

\date{\today}

\begin{abstract}
In this work, we study how to infer the orbit of a supermassive black hole binary (SMBHB) by time-dependent measurements with Very Long Baseline Interferometry (VLBI), such as the Event Horizon Telescope (EHT). Assuming a point-like luminosity image model, we show that with multiple years of observations by EHT, it is possible to recover the SMBHB orbital parameters -- eccentricity, (rescaled) semi-major axis, orbital frequency, and orbital angles -- from their time-varying visibilities even if the binaries orbital period is a few times longer than the duration of observation. Together with the future gravitational wave detections of resolved sources of SMBHBs with Pulsar Timing Array, and/or the detections of optical-band light curves, we will be able to further measure the individual mass of the binary, and also determine the Hubble constant if the total mass of the binary is measured through the light curves of the two black holes or measured by alternative methods. 

\end{abstract}


\section{Introduction}

Most galaxies harbor supermassive black holes in their centers. Binaries of supermassive black holes may form as a consequence of mergers of galaxies \citep{KormendyRichstone1995, Kauffmann2000, Volonteri2003, Ferrarese2005, KormendyHo2013, ColpiDotti2011}. Understanding the formation and evolution of supermassive black hole binaries (SMBHBs) are  essential to reveal the evolutionary histories of galaxies. 
The evolution of SMBHBs may be classified into several stages \citep{Begelman1980Nature}, depending on their separation and driving mechanisms. At their early stage with wide separations, the dynamical friction is capable of bringing the SMBHBs' separations down to order parsecs within cosmological timescales \citep[see, e.g., ][]{Callegari2011, Mayer2013, Dosopoulou2017}. At the separation of sub-parsec scales  (i.e. $< 0.01{\rm pc}$), gravitational wave emission is efficient to take away the energy and angular momentum of the SMBHB, so that they can merge within the Hubble timescale. As there is a gap between these two regimes for parsec-scale separations, it has been a long debate (the ``final parsec problem") whether and how SMBHBs migrate across the gap \citep{Begelman1980Nature, Colpi2014}. To overcome this final parsec problem, multiple mechanisms have been put forward to explain the efficient orbital damping, e.g., through the  interactions of SMBHBs with environmental  gas or stars in asymmetric nuclear potentials and on elongated orbits \citep[e.g., see the review by][]{Colpi2014}. To test these supposes, astrophysical observations over SMBHBs at sub-parsec separations are needed. 

Currently, the ongoing observations on SMBHBs include telescopes targeting electromagnetic signals within multiple frequency bands: radio, optical/infrared, X-ray, and also gravitational wave (GW) signals. In particular, the GWs emitted by SMBHBs at close separations are promising sources of pulsar timing arrays (PTAs) \citep{McLaughlin2013, Hobbs2013, PTA2016, PTA2020} at the frequency band of nano-Hz, or a period of order one year. In the future, the space-borne gravitational wave detector  LISA (Laser Interferometer Space Antenna) \citep{Amaro_etal2017} will be able to detect the merger signals of SMBHB coalescence, which  are the loudest and most energetic GW events in our universe. Currently, the observational evidence for SMBHBs is all conducted by direct or indirect electromagnetic observations. 

There are three known SMBHB systems found by direct electromagnetic imaging, which are identified as two distinct active galactic nuclei with projected separations of tens to thousands of parsec, in the radio, optical, and X-ray wavelengths \citep{Komossa2003ApJL, Rodriguez20060402, Fu2011}. 
While nearly all the sub-parsec SMBHB systems are spatially unresolved, their identification rely on indirect methods, such as the commonly-used approach based on the semi-periodicity variation,  including the emission-line dynamics \citep[e.g., ][]{Bogdanovic2009, ShenLoeb2010, Tsalmantza2011, Eracleous2012, Decarli2013, McKernan2013, Shen_etal2013ApJ, Liu2014ApJL, Liu_etal2016}, semi-periodic jet structures \citep[e.g., ][]{Begelman1980Nature, Conway1995}, semi-periodic light curves \citep[e.g., ][]{Graham2015, DOrazio:2015nature, Andjelka2019, Saade2020, Komossa2021, Komossa:2021exd}, tidal disruption event light curves \citep[e.g., ][]{Liu2009ApJL, StoneLoeb2011, Liufukun2014, Coughlin2017}, and orbital motion of an  unresolved radio core observed with very long baseline interferometry (VLBI) \citep[e.g., ][]{Sudou2003, D_Orazio_2018VLBI, Breiding2021}. Several SMBHB candidates have been selected out with these indirect methods \citep{Valtonen2008Natur, Bogdanovic2009, Liufukun2014, Graham2015, LiYR2017}. 

In this work, we focus on the radio-band observation, which has the best chance of spatially resolving SMBHBs. We discuss the question that, whether one can fully recover the orbit parameters of the binary based on the radio interferometry measurement. This problem is nontrivial as the orbital parameters contain the eccentricity, semi-major axis, orbital frequency, and various Euler orbital angles - they may contribute to the visibility function with various degrees of degeneracy. For a given observation time, multiple sources with different periods may be simultaneously monitored. It is then interesting to find whether the orbit tomography for systems with periods longer than the observation duration can still be successful. We will study systems in different parameter regimes to answer these questions.

Some of the resolved SMBHBs may have close separations such that their gravitational wave emission  are detectable by the pulsar timing arrays. If these ``golden" binary exist, we can combine the multi-messenger data to further determine the individual masses of the binary. Similarly,  if the information from other electromagnetic frequency  band is available, e.g., optical light curves, we may further use these golden binaries to independently measure the Hubble constant.


This paper is organized as follows. In Section~\ref{image_SMBHB}, we calculate the image and the visibility of SMBHB on the sky plane. In Section~\ref{SMBHB_likelihood_estimation}, in three representative examples of SMBHBs, we figure out the posterior distribution of their orbital parameters by doing the Markov-Chain Monte-Carlo simulations, and then we compare the ability of the constraining of SMBHB parameters for possible varying detection conditions. In Section~\ref{discuss}, we discuss the multi-messenger applications when combining the image detections proposed in this paper with the future GW detections by PTA (in subsection~\ref{multimessenger}), as well as the multi-frequency applications when combining the radio image detections with the detections from optical band light curves of the individuals (in subsection~\ref{multifrequency}). In the appendix, we give our mathematical proofs of the ways to break parameter degenerations. 

In this paper we use natural unit with $c=G=1$. 

\section{Imaging the supermassive black hole binary}
\label{image_SMBHB}

\begin{figure}[h] 
\centering
\includegraphics[height=7.7cm]{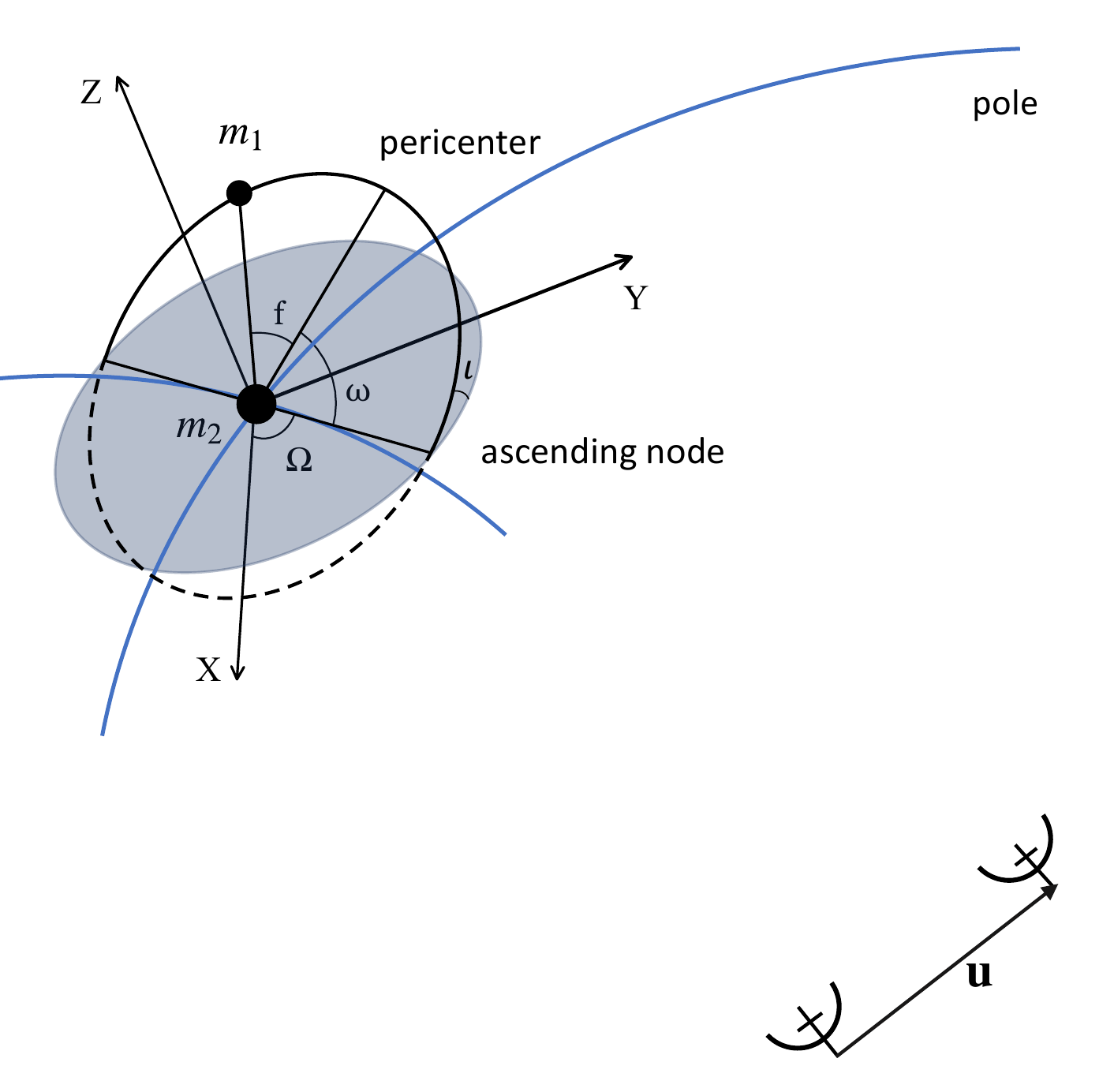}
\caption{The image and orbit of the supermassive black hole binary. The orbital angels are defined in the reference frame $(X, Y, Z)$ where $X$, $Y$ axises lying in the sky plane. Here, $\iota$ is the orbital inclination angle, $\omega$ is the periapsis, $\Omega$ is the angle of the longitude of ascending node, and $f$ is the phase of the individuals concerning the pericenter. The red-shifted mass of the individuals in the binary are $m_1$ and $m_2$.  }
\label{BBH_image}
\end{figure}

As SMBHs move within a gas-rich environment, electromagnetic radiations in various frequency bands may be sourced from locations  such as the circumbinary disk, circumsingle disks, possible jets, etc. It is a highly nontrivial task to model the emission in a given band as a function of accretion disk conditions and SMBHB orbital parameters, which requires  systematic numerical studies which are not currently available. As the first step to investigate the possibility of orbit tomography, 
 we adopt a simple analytical model, assume the emission from an SMBHB is described by two individual point-like luminosity functions given by 
\m \label{def_intensity}
I({\bf r})={I}_1 \delta ({{\bf r}}-{ {{\bf r}}_1}) +{I}_2 \delta ({{\bf r}} -{{{\bf r}}_2})\,,
\n 
where $I_1$ and $I_2$ are the intensities of the individuals, ${\bf r}$ is the sky position in radians, and ${\bf r_1}$ and ${\bf r_2}$ are the positions of the two components in the sky plane. This simple analytical model focuses on the emission in the vicinity of individual black holes and neglects emission from extended regions in the circumbinary disk, as well as possible variation of the luminosity function within orbital timescales. Our simplified image model describing SMBHB as two-point emitters moving along the eccentric and oblique orbit is illustrated in Figure \ref{BBH_image}. 

The complex visibility of the  SMBHB is defined to be the Fourier transform of their sky image, 
\m \label{define_visibility}
V({\bf u})=\int I({\bf r}) e^{- 2 \pi i {\bf u} \cdot {\bf r}}d^2 {\bf r}\,,
\n
where ${\bf u}$ is the vector baseline projected orthogonal to the line of sight and measured in wavelengths. 
The polar coordinate components for ${\bf r}$ and ${\bf u}$ in the sky plane are $(r, \phi)$ and $(u, \varphi)$. 
After plugging Eq.~(\ref{def_intensity}) back into Eq.~(\ref{define_visibility}), the visibility in Eq.~(\ref{define_visibility}) becomes
\m 
V({{\bf u}})=e^{-i 2 \pi {\bf u} \cdot {\bf r}_1}(1+ e^{i 2 \pi {\bf u} \cdot ({\bf r}_1-{\bf r}_2)}) \,.
\n
We shall focus on the visibility amplitude measurements in this study, which is
\m \label{abs_visibility}
|V({{\bf u}})|=\sqrt{ {I_1}^2+{I_2}^2+2 I_1 I_2 \cos{(2 \pi {\bf u}\cdot({\bf r}_1-{\bf r}_2))} }. 
\n
%
We define ${\bf R}$ as the projected vector of the binary separation vector ${\bf r}^{\prime}_{12}$ in the sky plane, and it is related to ${\bf r_1}-{\bf r_2}$ by the angular diameter distance $L$ as ${\bf R}=({\bf r_1}-{\bf r_2})L$. 
The separation vector ${\bf r}^{\prime}_{12}$ is determined by Keplerian orbit, with $|{\bf r}^{\prime}_{12}|={a (1-e^2)\over 1+e \cos f(t)}$, where $e$ and $a$ are the eccentricity and  semi-major axis of the binary orbit. The components of ${\bf r}^{\prime}_{12}$ in the reference frame $(X,Y,Z)$ can be obtained through the transformation laws of the Euler angels $(\Omega, \iota, \omega)$ by
\m 
{\bf r}^{\prime}_{12}&=&|{\bf r}^{\prime}_{12}|    \nonumber\\
&\times&
\left(
\begin{array}{c}
    \cos \Omega \cos (f(t)+\omega )-\cos \iota \sin \Omega \sin (f(t)+\omega ) \\
     \cos \iota \cos \Omega \sin (f(t)+\omega )+\sin \Omega \cos (f(t)+\omega ) \\
     \sin \iota \sin (f(t)+\omega ) \\
\end{array}
\right) ,   \nonumber\\
\n
therefore, 
\m \label{def_R}
{\bf R}&=&{a(1-e^2) \over 1+e \cos f(t)} \nonumber\\
&\times &
\left(
\begin{array}{c}
    \cos \Omega \cos (f(t)+\omega ) -\cos \iota \sin \Omega \sin (f(t)+\omega ) \\
     \cos \iota \cos \Omega \sin (f(t)+\omega )+\sin \Omega \cos (f(t)+\omega ) \\
\end{array}
\right) . \nonumber\\
\n

With the coordinate expressions, we now rewrite the visibility in Eq.~(\ref{abs_visibility}) as
\m \label{absV}
|V({{\bf u}})|&=&\sqrt{ {I_1}^2+{I_2}^2+2 I_1 I_2 \cos{(2 \pi {\bf u}\cdot{\bf R}/L)} } \nonumber\\
&=&\sqrt{ {I_1}^2+{I_2}^2+ 2 I_1 I_2 \cos{\Phi(t)} } \,, 
\n
where the phase $\Phi(t)$ in the cosine function is given by 
\m \label{def_Phi}
\Phi(t)&=&\frac{2 \pi (1-e^2) }{1+e \cos f(t)} {ua\over L} ( \cos \iota  \sin (\varphi \!-\! \Omega ) \sin (f(t)  \!+\! \omega ) \nonumber\\
&+& \cos (\varphi \!-\! \Omega ) \cos (f(t) \!+\! \omega ) ) \,. 
\n
The visibility amplitude is a function of time through $f(t)$, according to the following relations that describe motion within an eccentric Keplerian orbit:
\m \label{f_g}
\cos{f}&=&{\cos{g}-e\over 1-e \cos{g}} \,,\\
\tan{f\over 2}&=&\left({1+e\over1-e}\right)^{1/2}\tan{g\over 2}\,,\\
\label{g_t}
g-e \sin{g}&=&\omega_0 t\,,
\n
where $\omega_0=(m_1+ m_2)^{1/2}(a(1+z))^{-3/2}$ is the orbital frequency of the binary seen by the observer. Notice that the function $f(t)$ obtained from Equations~(\ref{f_g})-(\ref{g_t}) implicitly assumes a starting phase $f(0)=0$ at the starting time of observation. For generic initial conditions, we will need  an initial phase $f_0$ at the beginning of the observation. For the general case, the function $f(t)$ in Equation (\ref{def_Phi}) is replaced with $f(t+t_0)$, where $f(t_0)=f_0$.   

\section{Parameter estimation of the SMBHB orbit}
\label{SMBHB_likelihood_estimation}

To recover the orbital description for an SMBHB, there are nine unknown parameters to be determined by the visibility measurement,  including the ratio of $a/L$ (where $a$ is the proper semi-major axis and $L$ is the angular diameter distance), the (red-shifted) orbital frequency $\omega_0$, intensities $I_1$ and $I_2$, eccentricity $e$, inclination angle $\iota$, periapsis $\omega$, the longitude of ascending node $\Omega$, and initial phase $f_0$. In Appendix. \ref{appendix}, we provide a mathematical procedure to show how to obtain the value of these orbital parameters assuming perfect measurement without errors. Although realistic data always comes with measurement uncertainties, this mathematical procedure shows that there is no intrinsic degeneracy between different parameters that prevent the exercise of orbit tomography. 

Since the visibility is determined by the projected separation vector ${\bf R}$ of the binary, the projected orbital motion of the SMBHB on the celestial sphere can be traced by the measurements of the SMBHB visibility. If the visibilities observed by the two baselines ${\bf u}_1$ and ${\bf u}_2$ are $V_1(t)$ and $V_2 (t)$, 
then, from which observations, in principle, we could obtain 
\m \label{V_para_perp}
2 \pi {u} {R}_{1}/L&=&\Phi(\varphi \to {\varphi}_1)\,, \nonumber\\
 2 \pi {u} {R}_{2}/L&=&\Phi(\varphi \to {\varphi}_2)\,,
\n 
with $\varphi_{1,2}$ associated with the directions of the baselines, according to Eq.~(\ref{absV}). 
The information from two baselines is sufficient to determine the projected motion of the binary on the source plane, while additional baselines should provide better constraints on the orbit.
In general, assuming independent observations at different observing times and from  different baselines, the likelihood function can be written as:
\m \label{def_likelihood}
\mathcal{L} &=& \prod_m \prod_{n} {1\over \sqrt{2 \pi {\sigma_{m n}}^2}} e^{-{(V_{m n} - V_m (t_n) )^2 \over 2{\sigma_{m n}}^2 } }   \,,
\n
where $m$ is the index for baselines and $n$ is the index for the observation time  $t_n$, $V_m(t_n)$ is the observed visibility amplitude by the $m$th baseline at  time $t_n$, and $\sigma_{m n}$  is the expected measurement error bar of the corresponding data point.  For the sake of illustration, we assume two orthogonal baselines given by $\varphi_1=0$ and $\varphi_2=\pi/2$ for the examples presented below.
\begin{table}[h]
\caption{The parameters for the SMBH binaries are considered in the three examples.}
    \centering
    \begin{tabular}{ccccc}
\hline
\hline
example & $M$ ($\Msun$) & $a$ & $\theta (\mu {\rm as})$ & period (years) \\
\hline
1 & $1\times 10^9$ & $400M$ & $4$ & 10 \\
2 & $1\times 10^9$ & $550M$  & $5.5$ & 15 \\
3 & $2\times 10^9$ & $400M$  & $8$ & 20 \\
%
%
\hline
\end{tabular}
     \begin{tablenotes}
     \item {\bf Notes}:  The sources are taken to be at a distance of $1 {\rm Gpc}$ to earth (equivalent to an angular diameter distance of $L=1 {\rm Gpc}/(1+z)$, and a redshift $z=0.23$), and $\theta$ is the angular separations of these binaries.  
\end{tablenotes} 
\label{table:SMBHs} 
\end{table}

\begin{figure*}[h!] 
\centering 
\includegraphics[height=17.5cm]{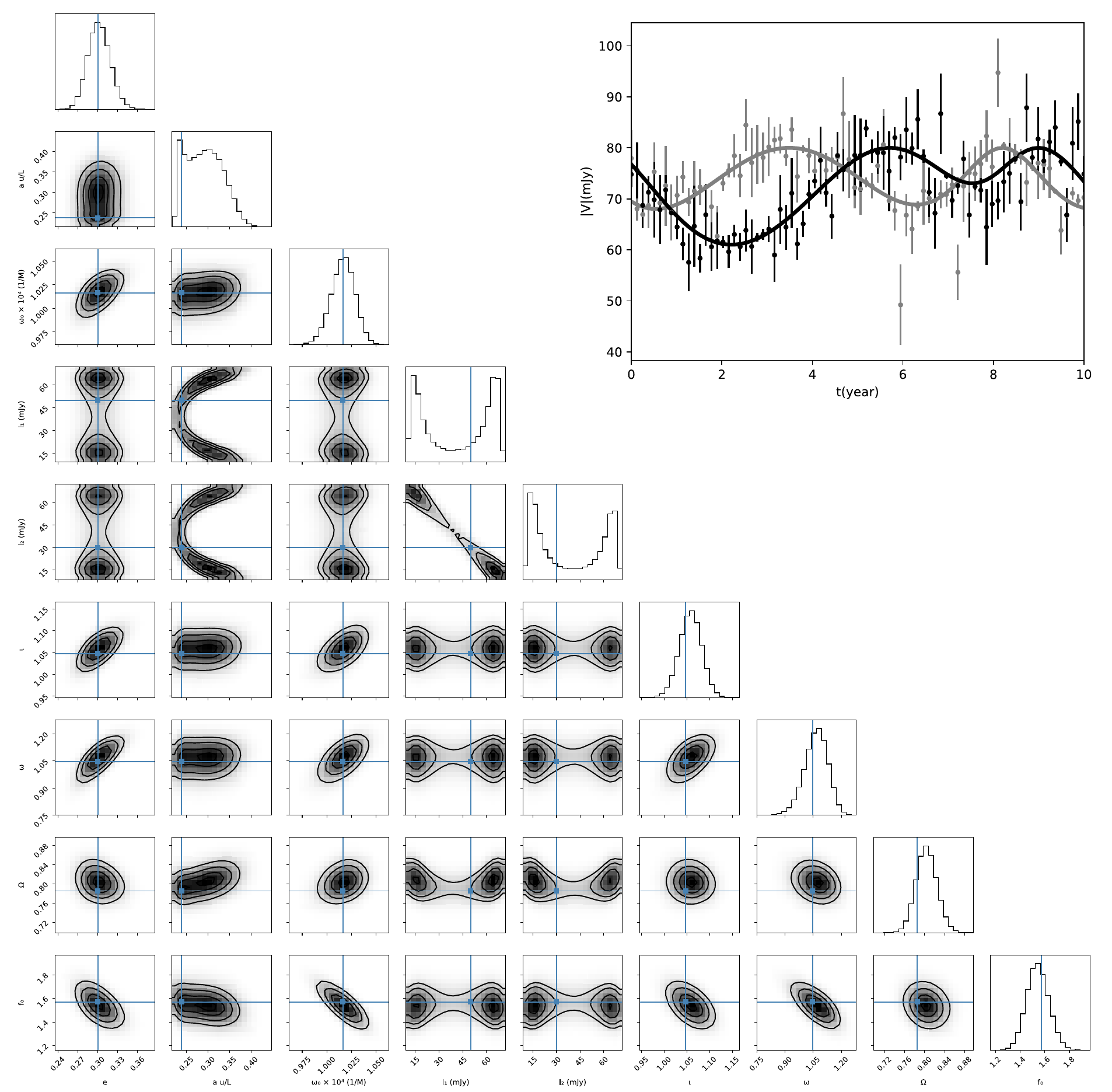} 
\caption{The estimation of the orbital parameters for the SMBHB is considered in our first example in Table~\ref{table:SMBHs}. The fitting of the parameters is conducted by sampling with Markov-Chain Monte-Carlo simulations. The underlying orbital parameters of the SMBHB are assumed to be $e=0.3$, $\iota=\pi/3$, $\Omega=\pi/4$, $\omega=\pi/3$, and $f_0=\pi/2$. The upper right plot shows the true values and the assumed error bars of the visibility function $|V|$ in this example: the curves represent the underlying value for $V$ (gray for $V_{1}$ and black for $V_{2}$),  and the points represent  the measurement data. The observation time is approximately equal to the orbital period of the SMBHB considered in this example.  }
\label{a400M_N80_eg1}
\end{figure*}

\begin{figure*}[t!] 
\centering 
\includegraphics[height=17.5cm]{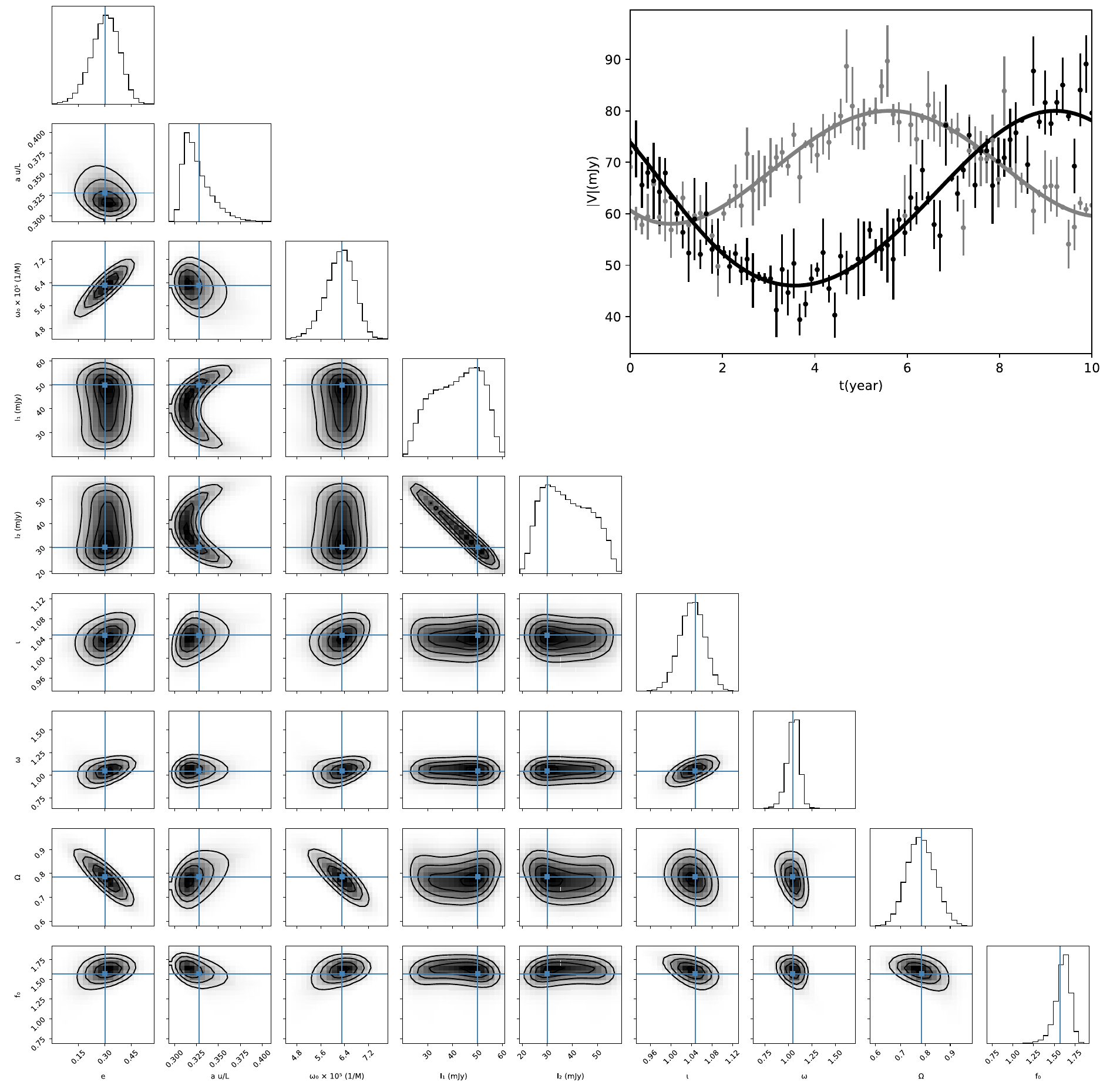} 
\caption{The estimation of the orbital parameters for the SMBHB is considered in our second example in Table~\ref{table:SMBHs}. The  orbital parameters (except for the orbital separation) are assumed to be the same as in Fig.~\ref{a400M_N80_eg1}. The upper right plot shows the true values and the  error bars of the visibility function $|V|$, where the curves represent  the true values (grey for $V_{1}$ and black for $V_{2}$), and the points represent the measurement data. The observation time is about $2/3$ of the orbital period of the SMBHB.    }
\label{a550M_N80_eg2}
\end{figure*}

\begin{figure*}[t!] 
\centering 
\includegraphics[height=17.5cm]{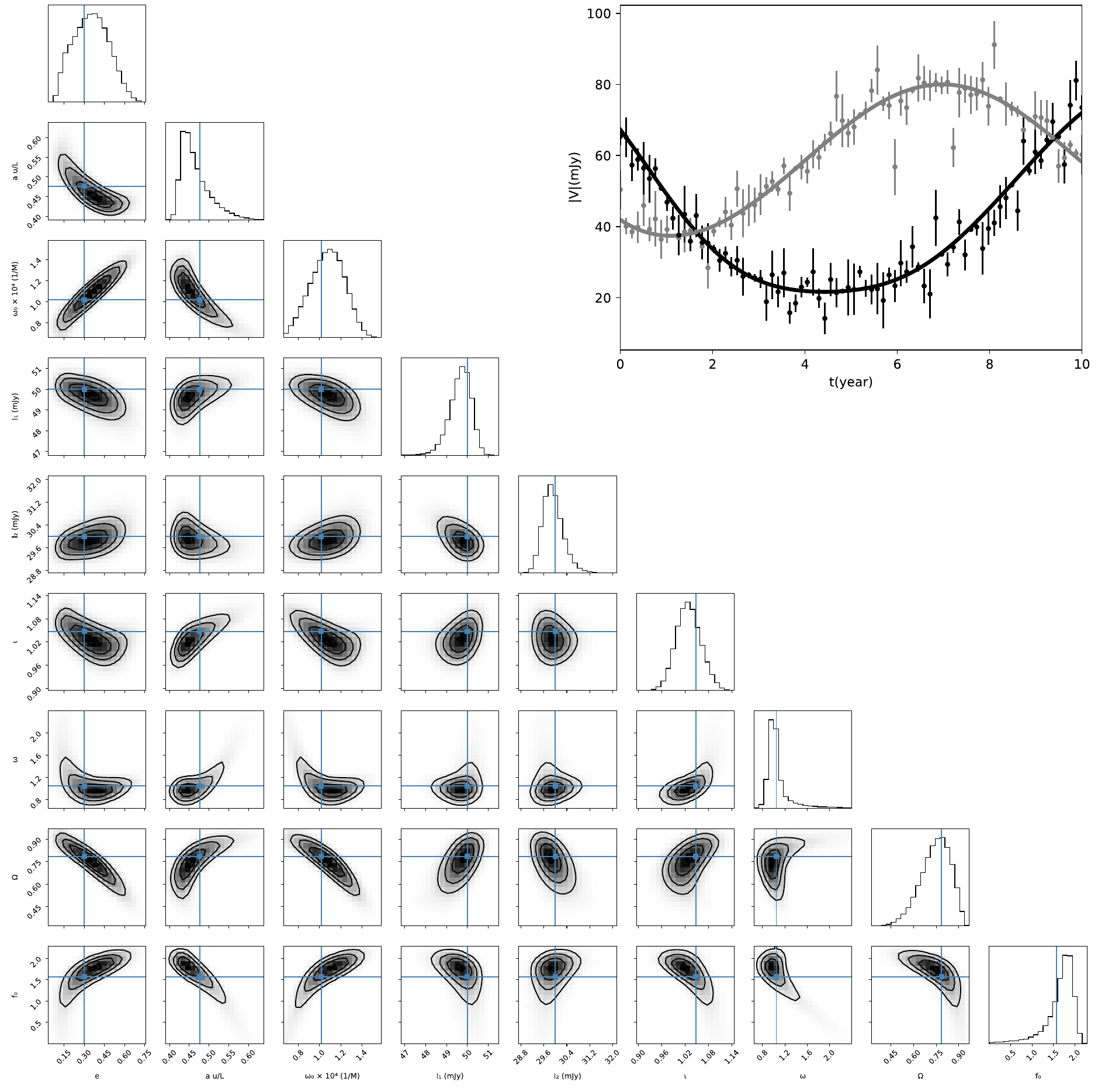} 
\caption{The estimation of the orbital parameters for the SMBHB is considered in our third example in Table~\ref{table:SMBHs}. The  orbital parameters (except for the ones displaced in the Table) are assumed to be the same as in Fig.~\ref{a400M_N80_eg1}.  The observation time is just one-half of the orbital period of the SMBHB.    }
\label{a400M_N80_eg3}
\end{figure*}

To illustrate the procedure to recover the orbital parameters, we consider three examples listed in Table~\ref{table:SMBHs}, indicating SMBHBs at 1Gpc distance, with an observed orbital period of 10, 15, and 20 years respectively. The second and the third examples represent SMBHBs with orbital periods larger than the observation period of EHT. The intensity of the individual black holes in the binary is set to be $I_1=50 {\rm mJy}$ and $I_2=30 {\rm mJy}$, which are the sample values taken from the low luminosity AGNs \citep{D_Orazio_2018VLBI}, resulting in a total intensity of $80 {\rm mJy}$. With this total intensity, we set the error bars of the modeled visibility for these sources to be of several ${\rm mJy}$, as is estimated according to the observations from \citep{Breiding2021VLBI}. The error bar $\sigma$ in each observation is sampled from a $[0, 8{\rm mJy}]$ uniform distribution times with a sample from the standard normal distribution (see Figures.~\ref{a400M_N80_eg1}-\ref{a400M_N80_eg3} for the details). The observation wavelength is assumed to  $\lambda=1{\rm mm}$. 
The two dimensionless baselines $u_{1}$ and $u_{2}$ measured by ${\rm mm}$ radio wavelength are taken to be $u_{1,2}=1.5 \times {\rm earth\ radius}/ \lambda$. 
For SMBHBs with greater separations, the orbital period will be too large to generate sufficient  variation in the visibility function within the observation period for us to recover their orbital parameters. For more compact SMBHBs, i.e., the ones with angular separations smaller than $2\mu {\rm as}$, the amplitude of variation of the visibility function will be too small to be resolved under the assumed condition of detection. The spatial resolution can be improved for visibility data with a higher signal-to-noise ratio or more arrays of baselines as is discussed in the rest part of this section. 

Given the Likelihood function, we apply the Markov-Chain Monte-Carlo method to obtain the posterior distributions of various orbital parameters, assuming flat priors. For the examples listed in Table~\ref{table:SMBHs}, the posterior distributions are shown in Figures~\ref{a400M_N80_eg1}-\ref{a400M_N80_eg3}, assuming ten years of observation time with VLBI.  During this observation period, we assume a uniform sampling rate of  8 times per year - equivalently 80 times in 10 years observation period (so that n ranges from 1 to 80 in Eq.~(\ref{def_likelihood})). 
The true evolution of the visibility functions and the assumed error bars of the measurement data  are shown in the upper right corners of the Figs.~\ref{a400M_N80_eg1}-\ref{a400M_N80_eg3}. In these examples, the minimal value of $|V|$  does not reach $|I_1-I_2| $,   as the phase factor $\Phi$  never reaches $\pi$ with the binary separation and distance assumed in Table.~\ref{table:SMBHs} (for all these cases $a/L$ is less than $1/u$). 

%

Based on the results shown the Figs.~\ref{a400M_N80_eg1} - \ref{a400M_N80_eg3}, we find that the orbital  parameters of the SMBHBs can be reasonably constrained with the assumed radio interferometry measurements. The fitting of the time-varying visibility is the key to recovering these parameters. 
By comparing the performance of orbit tomography for the three binaries, we conclude that: a. for a SMBHB in the example showed in Fig.~\ref{a400M_N80_eg1} which has orbital period no more than the observation time, i e., 10 years, the orbital frequency $\omega_0$ and eccentricity $e$ are better constrained which have an error bar within several percent, while the flux density is less constrained due to the smaller variation of visibility amplitude; b. for a moderate binary in Fig.~\ref{a550M_N80_eg2} which has orbital period beyond 10 years but within 20 years, the ratio $a/L$ is better constrained to within several percent, while the flux densities $I_1$ and $I_2$ are still indistinguishable; c. for a SMBHB shown in Fig.~\ref{a400M_N80_eg3} with larger orbital period (of order 20 years), the constraint over the orbital parameters are worse compared to the previous two cases except for the flux densities, since the variation of visibility amplitude is larger; d. both the inclination angles $\iota$ in these three examples are well constrained to be percent level. 

\begin{figure}[t!] 
\centering 
\includegraphics[height=16cm]{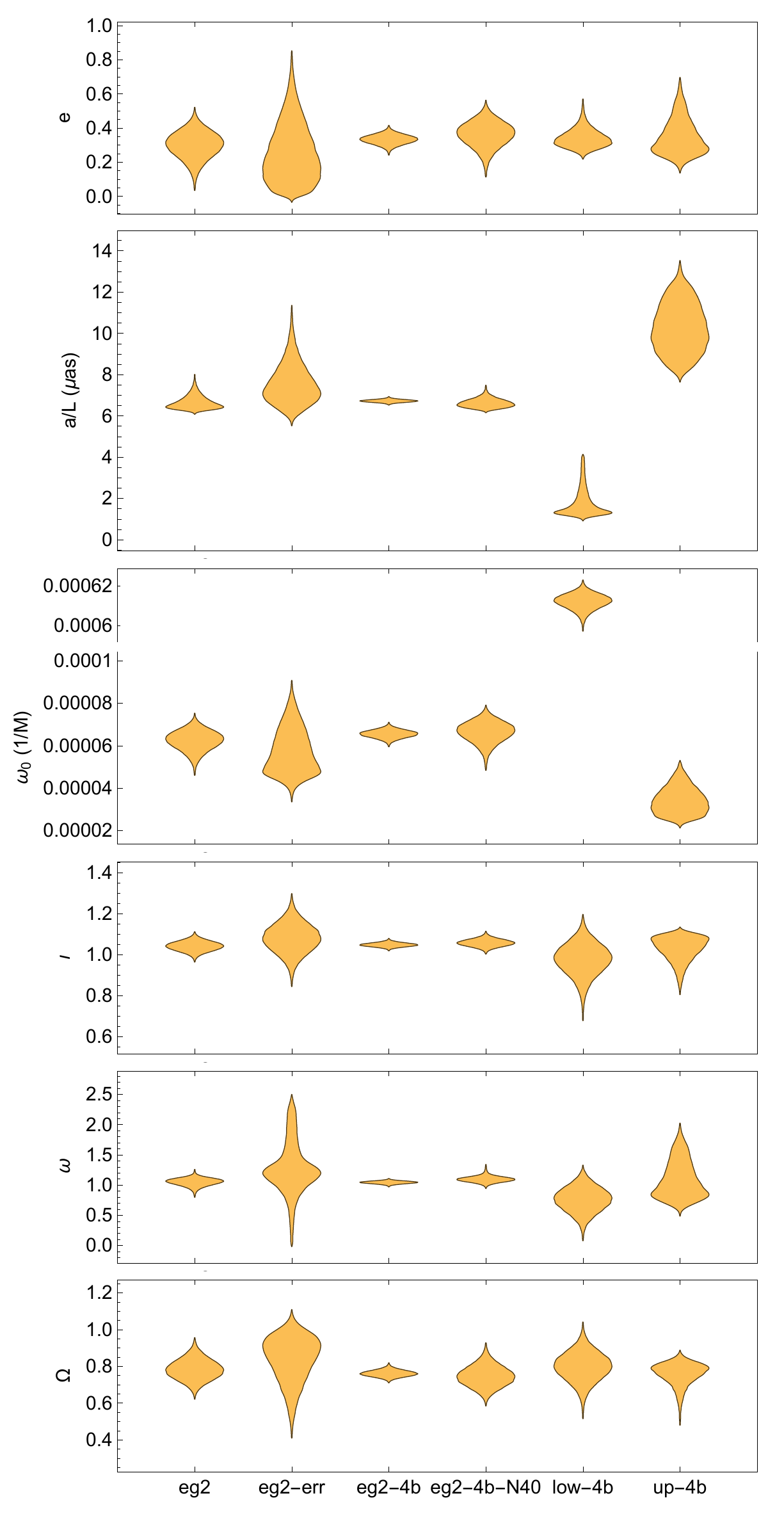} 
\caption{The posterior distributions of the SMBHB orbital parameters assuming various source configurations and/or detection conditions with VLBI interferometry. The first column uses the the same parameters as the second example in Table~\ref{table:SMBHs} (i e., Figure~\ref{a550M_N80_eg2}). The second column  assumes the same binary parameter as the first column, but with a different noise model that  has a multiplicative dependence on $|V|$. The third column represents the case with four baselines oriented in different angles.
The fourth column assumes 40 rounds of observation in total (4 times per year) which is half as frequent as other cases.
 The fifth and sixth columns represent the cases with smaller and larger binary separations respectively.      }
\label{param_violin_plot}
\end{figure}

We further explore the ability to constrain SMBHB parameters with a different number of baselines,  error models, observation schedules,  and orbital separations. The dependence of the posterior distributions of the binary parameters on these variations is presented in Figure~\ref{param_violin_plot}.  In the four baselines case, we assume the directions of the baselines are at $\varphi=0, \pi/4, \pi/2, 3\pi/4$ respectively.  The corresponding distributions (see the third column with the label ``eg2-4b", which represents the second example in Table~\ref{table:SMBHs} with four baselines) are significantly narrower than those of the two-base case (``eg2"). 
The improvement in accuracy can be attributed to the additional information brought by  the additional baselines. 
By comparing the first and the fourth column we could see, the constraining ability of the parameters of SMBHB in the second example with two baselines and 8 times observation frequency per year are almost the same as the results constrained by four baselines while with 4 times observation frequency per year (``eg2-4b-N40"). 
In a separate example ( the second column in Figure~\ref{param_violin_plot} with label ``eg2-err"),  we test the performance of a different visibility error prescription, with ${{\sigma}^{\prime}}^2=(0.5 \sigma)^2+ (0.1 |V|)^2$ (where $\sigma$ is the error bar considered in Figures~\ref{a400M_N80_eg1}-\ref{a400M_N80_eg3}), so that the error also increases if the expected value of $V$ is larger. As the noise fluctuation for the visibility data is generally greater in this prescription, the constraints on the orbital parameters become worse as expected. 
A similar trend is observed as we increase the waiting time between observations shown in ``eg2-4b-N40". 
In the last two examples, we still assume an SMBH binary  with a total mass of $M=10^9\Msun$ at distance $1{\rm Gpc}$, four baselines, and the same  error bar model as considered in Figures~\ref{a400M_N80_eg1}-\ref{a400M_N80_eg3}). The binary separations are chosen to be  $1.2 \mu{\rm  as}$ (or 1.6 years in period) and $8.4 \mu{\rm as}$ (or 30 years) respectively.  In the lower separation case, the binary period is much smaller than the observation period. The time-dependent visibility data contains multiple oscillation cycles, but oscillation amplitude is small because of the smaller $a/L$. In this case, the orbital frequency  has the best relative precision among all the parameters.
In the larger separation case, the binary period is three times of the observation period. This means that the visibility measurement only lasts for a fraction of an oscillation cycle. The corresponding measurement accuracy of orbital frequency is much worse than the low separation case.  The orbital angle measurement uncertainties are both significantly larger than the ``eg2-4b" case with modest orbital separation.

For much more compact binaries, i.e., the ones detectable by the successor of LISA: AMIGO (Advanced Millihertz Gravitational-wave Observatory) \citep{Baibhav:2019rsa}, space-based VLBI is required to provide sufficient angular resolution to recover the binary orbit or even the final black hole. In this case, a multi-messenger test of General Relativity may be performed, as discussed in  \citep{Yang:2021zqy}. 

At this point, it is also worth thinking about how to claim a detection based on the visibility measurement. In principle, if we have multiple (more than two) baselines and the inferred binary parameters are consistent with each other assuming a different combination of baseline, this comparison provides a good indication that the underlying source is an SMBHB. Similar tests can also be performed at different wavelengths within the radio band. However, it will be difficult to distinguish the binary scenario from the case of two arbitrarily moving blobs within the accretion disk, if the binary period is significantly longer than the observation period. It requires a more detailed study, along with better binary-disk emission models, to assess the appropriate binary parameter range that allows successful model selection. 

\section{Multi-messenger/band detection} 
\label{discuss}

In addition to radio interferometry observations, some of the SMBHBs may be observed in other frequency bands, such as optical/infrared and X-ray. They may also be resolved by PTAs for a suitable range of parameters. We shall discuss two examples to illustrate what further information can be obtained from  multi-messenger/multi-band observations. 

\subsection{Multi-messenger observation with gravitational waves}
\label{multimessenger}
Both extreme-mass-ratio inspirals and SMBHBs are promising sources for multi-messenger detection with gravitational waves. Extreme-mass-ratio inspirals are mainly observed by LISA in their last stages of the inspiral/merger process, with the GW measurement providing the orbit information and  radio signals coming from the common accretion disk and/or the jet \citep{Pan:2021ksp,Pan:2021oob}. On the other hand, SMBHBs resolvable by ground-based VLBI should have much wider separation and much lower frequency. The SMBHBs emitting GWs at the PTA band should have negligible frequency evolution in the timescale of years so that the GW alone is insufficient to infer the orbit. Without the chirp signal as commonly seen for ground-based detection, one can only measure an overall amplitude $A = {\mathcal{M}_c}^{5/3}/D_L$ (where $\mathcal{M}_c$ is the red-shifted chirp mass and $D_L$ is the luminosity distance) coupled with inclination and polarization  phase angles. Assuming a joint measurement with radio VLBI and PTA, the luminosity distance can be inferred from the redshift of the host galaxy of the SMBHB, and the orbital angles can be measured with VLBI through the varying visibility of the SMBHB images, as is discussed in Section~\ref{SMBHB_likelihood_estimation} (see Figures~\ref{a400M_N80_eg1}-\ref{a400M_N80_eg3}), which breaks the degeneracy between the chirp mass, luminosity distance, and the orbital angles in the strain amplitude of GW.  
We may obtain the chirp mass of the binary up to measurement uncertainty. 
Suppose that we  detect the GWs of an SMBHBs through PTA with a certain single-to-noise ratio (SNR), the uncertainty of the amplitude $A$ of the system can be estimated directly through the Fisher matrix as follows:
\begin{align} \label{chirp_mass_error}
{\Delta A}^2  = \frac{1}{\langle h,h \rangle} = \frac{A^2}{A^2\langle h,h \rangle} \sim {A^2 \over {\rm SNR}^2}\,,
\end{align}
where $h$ is the waveform of SMBHB, and $\langle,\rangle$ defines the inner product used in the Fisher matrix. Assuming the measurement uncertainty of the redshift and orbital angles are  smaller than the relative uncertainty of the amplitude (i.e., Figure.~\ref{a400M_N80_eg1}-\ref{a400M_N80_eg3} ), the relative uncertainty of the chirp mass would be similar to that of the amplitude. 
For simplicity, the posterior of the chirp mass $M_c$ is approximated by a Gaussian distribution 
\m 
P(\mathcal{M}_c)={1\over \sigma_{c} \sqrt{2 \pi}} e^{-{1\over 2} ({\mathcal{M}_c-\mathcal{M}_{c0} \over \sigma_{c} })^2}
\n
where $\mathcal{M}_{c0}$ is the truth value of the chirp mass, and $\sigma_c \approx \mathcal{M}_{c0}/{\rm SNR}$ is obtained from Eq.~(\ref{chirp_mass_error}). 

On the other hand, through the measurement with EHT,  the posterior distribution of the (red-shifted) total mass $\mathcal{M}$ of the binary could be extracted from the joint posterior distribution of $\omega_0$ and $a$ (see Fig.~\ref{a400M_N80_eg1}-\ref{a400M_N80_eg3}), by the Kepler's law:$\mathcal{M} = {\omega_0}^2 (a(1+z))^3$, with the angular diameter distance inferred from the redshift. 

\begin{figure}[h] 
\centering 
\includegraphics[height=5cm]{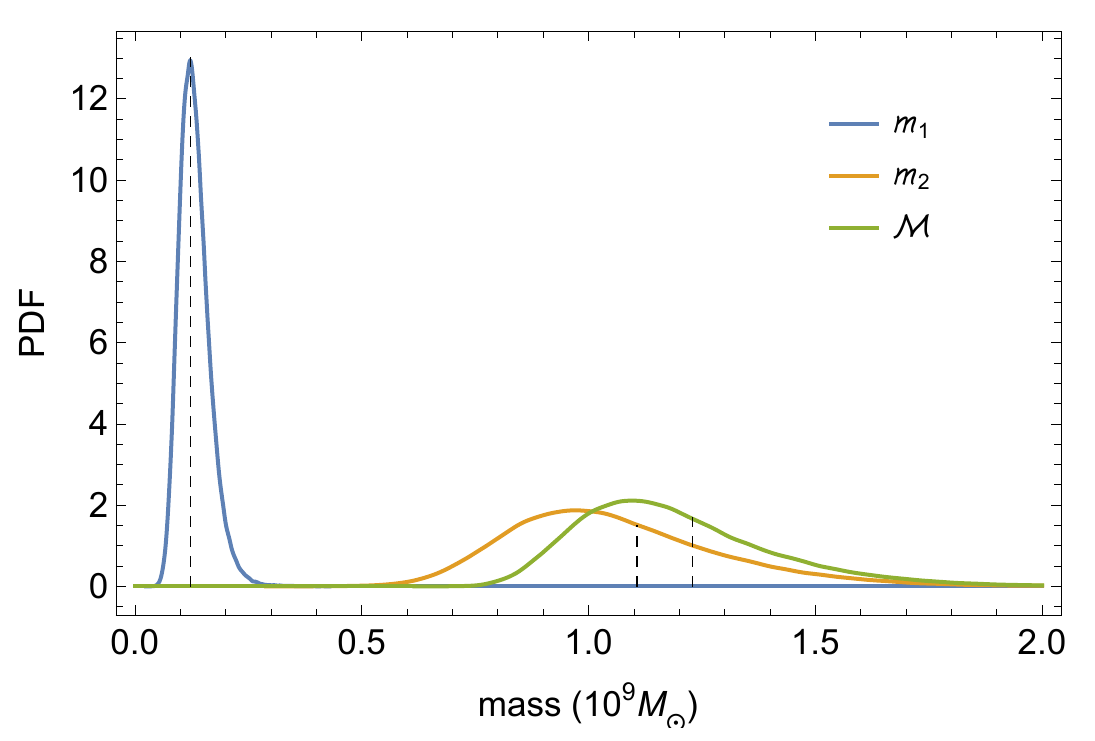} 
\caption{ The probability density function (PDF) for the individual masses of an SMBHB in Fig.~\ref{a550M_N80_eg2}, assuming the binary has a mass ratio of $9:1$. 
The green line is the posterior $P(\mathcal{M})$ obtained from Fig.~\ref{a550M_N80_eg2}. The vertical lines are the corresponding true values. 
   }
\label{mass_posterior_eg1}
\end{figure}

\begin{figure}[h] 
\centering 
\includegraphics[height=5cm]{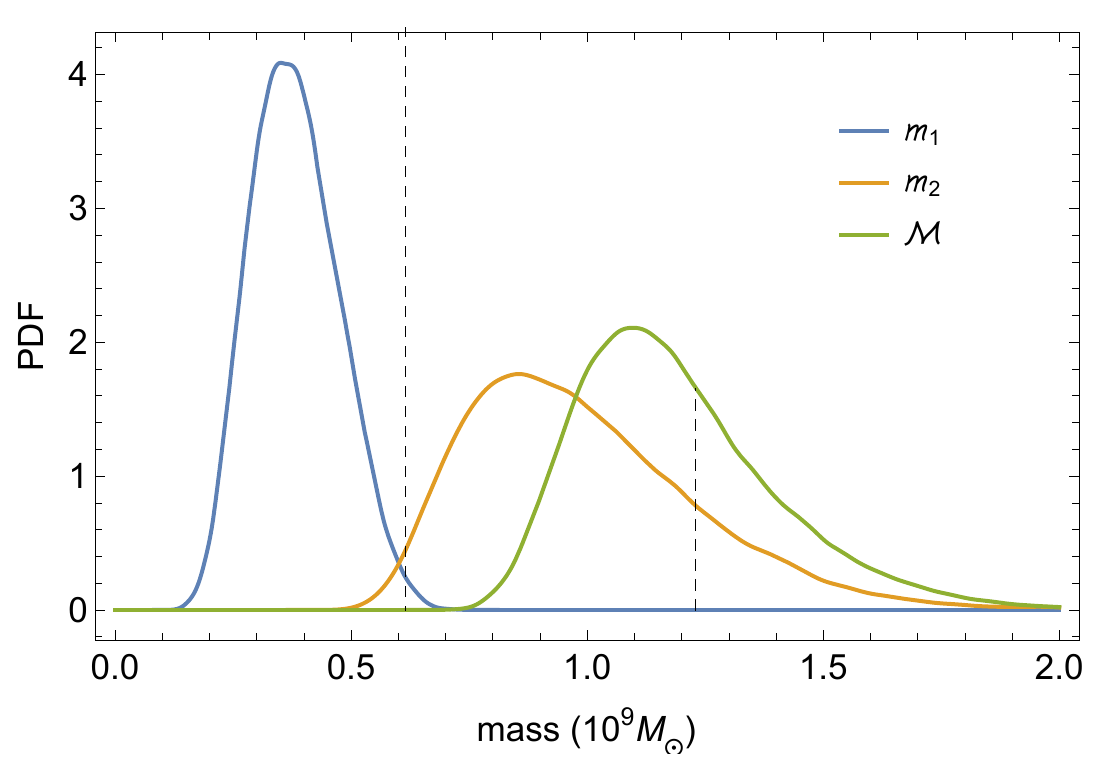} 
\caption{ The probability density function (PDF) for the individual masses of an SMBHB in Fig.~\ref{a550M_N80_eg2}, but for a binary with equal mass. The poor resolution of the individual mass is due to the errors from the measurements of $\mathcal{M}$ and ${\mathcal{M}}_c$. 
   }
\label{mass_posterior_eg2}
\end{figure}

Provided with the posterior distribution $P(\mathcal{M}_c)$ and $P(\omega_0,a)$, and the fact that $\mathcal{M}=m_1+m_2$ and $\mathcal{M}_c ={{m_1}^{3/5}{m_2}^{3/5}/{\mathcal{M}}^{1/5} }$, we will try to  recover the distributions of  individual mass $m_1$ and $m_2$, with $m_1$ defined to be the less massive mass. First we sample the points in $a, \omega_0$ and use them to compute the total mass $\mathcal{M}$. After that we sample the points in chirp mass according to its distribution $P(\mathcal{M}_c)$, and then compute individual masses
according to $m_1= {1\over 2}\mathcal{M}-{1\over 2}\sqrt{ {\mathcal{M}}^2 - 4 {\mathcal{M}}^{1/3} {\mathcal{M}_c}^{5/3} }$ and $m_2={1\over 2}\mathcal{M}+{1\over 2}\sqrt{ {\mathcal{M}}^2 - 4 {\mathcal{M}}^{1/3} {\mathcal{M}_c}^{5/3} }$, from which  we remove the samples giving complex numbers for $m_1$ and $m_2$.  The statistical distributions of all sampling points give the probability densities $P(m_1)$ and $P(m_2)$.  
For the system discussed in the example of Fig.~\ref{a550M_N80_eg2}, the results of the posterior distribution of $m_1$ and $m_2$ for a SMBHB has $SNR=10$ of PTA GW detection are shown in Fig.~\ref{mass_posterior_eg1} and Fig.~\ref{mass_posterior_eg2}, assuming different underlying mass ratio. 
%
%

The underlying mass ratio assumed in  Fig.~\ref{mass_posterior_eg1} and Fig.~\ref{mass_posterior_eg2} are $9:1$ and $1:1$ respectively, with the same radio measurement result shown in  Fig.~\ref{a550M_N80_eg2}. These assumptions may not be physical as one may expect the luminosity ratio to be correlated with the mass ratio. Nevertheless, for various mass ratios, the distributions for each component masses can be successfully constructed with the joint measurement, which agrees with the underlying injected values within $1-2 \sigma$. We also find that it is easier to separate out $m_1$ and $m_2$ in the first example, as expected.

\subsection{Multi-band observation to determine the Hubble constant}
\label{multifrequency}
Since SMBHBs may be observed simultaneously in different frequency bands of electromagnetic waves, it is instructive to discuss multi-band measurements in this context. Especially, the periodic light curve in the optical band was found in the system PG 1302–102 which is explained as the relativistic Doppler boost modulation on the flux densities of the SMBHB individuals \citep{Graham2015, DOrazio:2015nature}. 
And the periodic variability arising from relativistic Doppler boost is found to be a promising electromagnetic signature to connect with GW detections \citep{Charisi:2021dwc}. 
In this section, we consider the scenario that the orbital velocities of  individual black holes in the binary are measured using the modulation of flux densities, generated by the relativistic boost. If a black hole has a velocity of $v$ and a rest frame flux density of $F^0_{\mu}$, then the variation of the observed flux density $F_{\mu}$ assuming a general Keplerian orbital motion is 
\m
{|\Delta F_{\mu}|\over F_{\mu}}=(3-\alpha)v \big[ e \cos{\omega}+\cos{\left( f(t+t_0) + \omega \right)} \big] \sin{\iota}\,, \nonumber \\
\n
 where $v=m_i {\mathcal{M}}^{-1/2}  (a(1+z)(1-e^2))^{-1/2}$ (i=1, 2), and $\alpha$ is the exponent of the power-law that best describes the spectrum in the frequency region of interest. It is usually assumed to be $\alpha= 1.1$ as a good proxy for the optical V band \citep{DOrazio:2015nature, Dotti:2021bjm}. Here, the eccentricity $e$, inclination angle $\iota$, periapsis $\omega$, and  the phase angle $f_0$ may all be measured through the parameter estimation using the time-dependent visibility, as discussed in Sec.~\ref{SMBHB_likelihood_estimation}. In addition, measuring the optical light curves gives the instantaneous $v$. The black hole that has a lower mass with higher velocity will have a larger relativistic boost of the flux density.  
 
 The measurement of $v$ has two advantages when combined with the radio interferometry measurements. Firstly, based on the velocity data we can compute the posterior distribution of $m_i {\mathcal{M}}^{-1/2}$, which further determine the probability density functions of individual masses $m_1$ and $m_2$. Secondly, 
 as the angular separation   of the SMBHB is directly measured through radio interferometry, one can  determine the value of the angular diameter distance if the physical separation of the SMBHB is known. 
 Since $m_1, m_2$ can be inferred from the optical light curves, and both the orbital frequency and the host galaxy redshift $z$ are known, we can determine the physical separation $a$ by using the relation $\omega_0={\mathcal{M}}^{1/2}(a(1+z))^{-3/2}$. As a result, the angular diameter distance $L$  can be determined. This can serve as an independent approach to measure
the Hubble constant.

In addition to the optical light curve measurement, there are also alternative ways to measure the  mass of black holes within SMBHB, for example using the dynamical mass measurements or the relation between the SMBH mass and its host galaxy properties \citep[see, e.g., ][]{Peterson2014, Schutz:2015pza}. Provided with  mass measured from these alternative methods, we should also be able to determine the Hubble constant, similarly as mentioned above. One difference is that we can in principle detect the optical light curves at cosmological distances (say, about $1{\rm Gpc}$ \citep{Graham2015, DOrazio:2015nature, Valtonen2008Natur, Komossa2021}), while these alternative methods only resolve close sources (within about $L \sim 100 {\rm Mpc}$ \citep{Schutz:2015pza}).  

\section{Conclusion}

Assuming a point-emitter luminosity distribution, we have shown that the orbital parameters of an SMBHB can be recovered with time-dependent radio VLBI measurements. The orbit tomography is still possible if the observation period is a few times shorter than the period of the binary system.  If additional measurement with GWs (using PTA) and/or electromagnetic signals in other frequency bands are available, the component masses of these ``golden binaries" may be separately determined, and the joint multi-band observation may be used to measure the Hubble constant. Exploring the science potential of these golden SMBHBs is  likely a fruitful direction for the next-generation EHT.

In reality, the radio emission may come from not only the circumsingle disks around individual black holes but also the circumbinary disk region or even possible jets. As the tidal steam feeding gas onto the circumsingle disks can vary based on the orbital phase, the emission from the vicinity of the individual black holes may also have nontrivial time dependence in orbital timescales.
Therefore magneto-hydrodynamics simulations with SMBHBs moving in accretion disks are necessary to fully characterize the electromagnetic signals \citep[see, e.g., ][]{Gold:2013zma, Gold:2019nqg, Paschalidis:2021ntt}. Similar tasks are being carried out for binaries with smaller separations \citep[see, e.g., ][]{Farris:2012ux, Gold:2014dta} so that they fall into the detection band of LISA for multi-messenger observations. Once the radio emission is properly understood based on the systematic studies of numerical simulation, one may revisit the orbit tomography problem given all the environmental uncertainties from the accretion flow. The orbit tomography may no longer have a clear answer as shown in this work, but as the time-dependent visibility likely contains rich information about the accretion flow, the new inverse procedure may help us to constrain disk properties in addition to orbital parameters. 

\section*{Acknowledgements}
We acknowledge the use of the HPC Cluster of the National Supercomputing Center in Beijing. 
We would like to thank Luis C. Ho, Roman Gold, Zhenwei Lyu, Weiwei Xu, and Key Wang for useful discussions. 
Y. F. is supported by the National Science Foundation of China (NSFC) Grant No. 11721303, 
and the fellowship of China Postdoctoral Science Foundation No. 2021M690228. 
 H. Y. is supported by the Natural Sciences and
Engineering Research Council of Canada and in part by
Perimeter Institute for Theoretical Physics. Research at
Perimeter Institute is supported in part by the Government of
Canada through the Department of Innovation, Science and
Economic Development Canada and by the Province of Ontario through the Ministry of Colleges and Universities.

\section*{Appendix: recover the orbital parameters assuming perfect detection}\label{appendix}
In this section, we outline a mathematical procedure of resolving the orbital parameters $I_1$, $I_2$, $\omega_0$, $e$, $\iota$, $\omega$, $\Omega$, and $a/L$, assuming a perfect observation.  
Firstly, for a SMBHB which has a relatively large separation such that $\cos{\Phi(t)}$ could reach to the full range of $\[-1, 1\]$ (which is roughly equivalent to $a u/L>1$),  we may compute the vaules of $I_1$ and $I_2$ through the maximal and minimal values of $|V|$. 

\begin{figure}[h] 
\centering
\includegraphics[height=5cm]{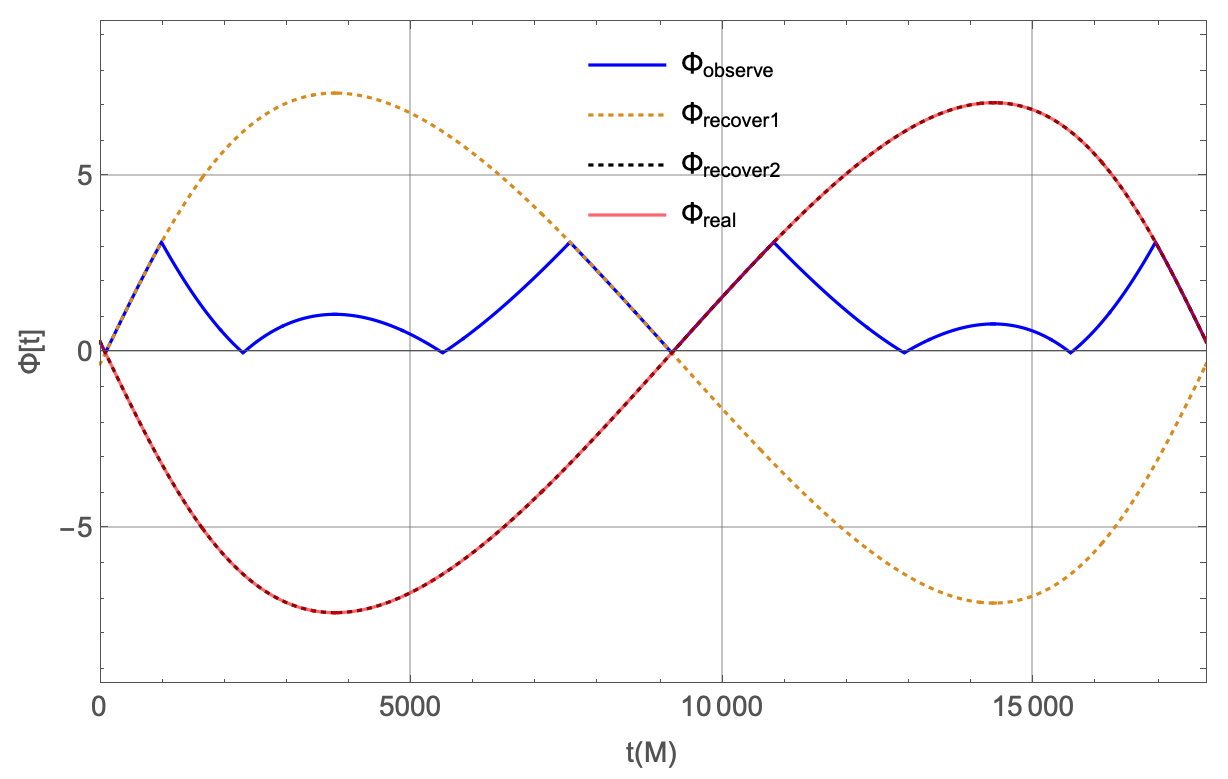} 
\caption{The case of SMBHB image visibility when the separation is large enough such that $\cos{\Phi(t)}$ could reach to the full the range of $\[-1, 1\]$, and the reconstruct of the function $\Phi(t)$. }
\label{recover_Phi}
\end{figure}

Secondly, we could recover the phase $\Phi(t)$ in Equation~(\ref{absV}) from the observed visibility $V(t)$ through
\m \label{determe_Phi}
{\Phi(t)}=\arccos{\left [ {|V(t)|-{I_1}^2-{I_2}^2 \over 2 I_1 I_2} \right ] }\,. 
\n
From Equations~(\ref{def_Phi})-(\ref{g_t}) we notice that $\Phi$ is a smooth function of $t$, so that we can obtain $\Phi(t)$ by solving Equation~(\ref{determe_Phi}) with the continuity condition of the derivative, up to a sign of $\pm$. In Figure~(\ref{recover_Phi}), we show an example of $\Phi(t)$ obtained using Equation~(\ref{determe_Phi}) ($\Phi_{observe}$), the recovered $\Phi$ ($\Phi_{recover1}=-\Phi_{recover2}$) and the real $\Phi$ ($\Phi_{real}$). 

Thirdly, we rewrite $\Phi(t)$ in Equation~(\ref{def_Phi}) as
\begin{widetext}
\m
\Phi(t)&=&2 \pi u (1-e^2){a\over L} \{ [ \cos\iota \sin \omega  \sin (\varphi -\Omega )+\cos \omega  \cos (\varphi -\Omega ) ] \frac{\cos f(t)}{1+e \cos f(t)} \nonumber\\
  &+&[ \cos \iota \cos \omega  \sin (\varphi -\Omega )-\sin \omega \cos (\varphi -\Omega )] \frac{\sin f(t)}{1+e \cos f(t)} \}\,, \nonumber\\
\n
and using the fact that (see, e.g., Equation~(4.87), (4.88) in \citep{Maggiore2007})
\m
\frac{(1-e^2)\cos f(t)}{1+e \cos f(t)}=-{3\over 2}e+\sum^{\infty}_{1} {1\over n}(J_{n-1}(ne)-J_{n+1}(ne)) \cos{n \omega_0 t}\,,
\n
and 
\m
\frac{(1-e^2)\sin f(t)}{1+e \cos f(t)}=\sum^{\infty}_{1} {\sqrt{1-e^2}\over n}(J_{n-1}(ne)+J_{n+1}(ne)) \sin{n \omega_0 t}\,,
\n
we  Fourier decompose $\Phi(t)$ as follows, 
\m \label{Fourier_V}
 \Phi(t)&=&2 \pi u {a\over L} \{ C_1 \sum^{\infty}_{0} a_n \cos{n \omega_0 t} + C_2 \sum^{\infty}_{1} b_n \sin{n \omega_0 t} \}
  \n
where
\m
C_1&=& \cos\iota \sin \omega  \sin (\varphi -\Omega )+\cos \omega  \cos (\varphi -\Omega )\,, \ 
C_2=\cos \iota \cos \omega  \sin (\varphi -\Omega )-\sin \omega \cos (\varphi -\Omega )\,,  \nonumber\\
a_0&=&-{3\over 2}e\,, \nonumber\\
a_n&=&{1\over n}(J_{n-1}(ne)-J_{n+1}(ne))\,, for\, n=1,2,... \nonumber\\
b_n&=&{\sqrt{1-e^2}\over n}(J_{n-1}(ne)+J_{n+1}(ne))\,, for\, n=1,2,... \nonumber\\
\n
Consider the observing time staring from an arbitrary initial phase $f_0=f(t_0)$, the phase function in Equation~(\ref{Fourier_V}) is 
\m \label{Fourier_V_f0}
 \Phi(t)&=&2 \pi u {a\over L} \big[ C_1 \sum^{\infty}_{0} a_n \cos{n \omega_0 (t+t_0)} + C_2 \sum^{\infty}_{1} b_n \sin{n \omega_0 (t+t_0)} \big] \nonumber\\
 &=& 2 \pi u {a\over L} \big[ C_1 a_0 + \sum^{\infty}_{1} \left(C_1 a_n \cos{n \omega_0 t_0} + C_2 b_n \sin{n \omega_0 t_0} \right) \cos{n \omega_0 t}  
 + \sum^{\infty}_{1} (C_2 b_n \cos{n \omega_0 t_0} - C_1 a_n \sin{n \omega_0 t_0} )\sin{n \omega_0 t } \big] \nonumber\\
  \n
\end{widetext}

With the above preparation, we now recover orbital parameters as follows:

{\bf{Orbital frequency $\omega_0$.}} We obtain $\omega_0$ through the variation period of the phase $\phi(t)$ of ${\bf R}$. Assuming $\Phi_{\parallel}$ and $\Phi_{\perp}$ are two phase function of two visibilities observed from two orthogonal baselines ${\bf u}_{\parallel}$ and ${\bf u}_{\perp}$, then the phase $\phi(t)$ of ${\bf R}$ could be obtained by 
\m
\phi(t)={\rm arg}\left[ {\Phi_{\parallel} \over {\Phi_{\parallel}}^2 +{\Phi_{\perp}}^2}  + i {\Phi_{\perp} \over {\Phi_{\parallel}}^2 +{\Phi_{\perp}}^2}  \right]\,.
\n
The time render $\phi$ to range a full circle is the orbital period $2\pi/\omega_0$, which gives $\omega_0$. 
%


{\bf Eccentricity $e$ and initial phase $f_0$.} We  integrate $\Phi(t)$ in Equation~(\ref{Fourier_V_f0}) by
\m \label{def_C_n}
{\mathcal{C}}_n=\int^{2\pi/\omega_0}_{0}\Phi(t) \cos{n \omega_0 t} dt&=&2 \pi u {a\over L} {\pi\over \omega_0} (C_1 a_n \cos{n \omega_0 t_0}  \nonumber\\
&+& C_2 b_n \sin{n \omega_0 t_0} ) \,,
\n
and
\m \label{def_S_n}
{\mathcal{S}}_n=\int^{2\pi/\omega_0}_{0}\Phi(t) \sin{n \omega_0 t} dt&=&2 \pi u {a\over L} {\pi\over \omega_0} (C_2 b_n \cos{n \omega_0 t_0}  \nonumber\\
&-& C_1 a_n \sin{n \omega_0 t_0} ) \,.
\n
Combining Equation~(\ref{def_C_n}) and Equation~(\ref{def_S_n}), we have 
\m \label{def_C1_n}
C_1 &=&2 {\pi}^2 {a u\over L \omega_0} {{\mathcal{C}}_n \cos{n \omega_0 t_0} - {\mathcal{S}}_n \sin{n \omega_0 t_0} \over a_n} \,, \\
\label{def_C2_n}
C_2 &=&2 {\pi}^2 {a u\over L \omega_0} {({\mathcal{C}}_n + {\mathcal{S}}_n \cot{n \omega_0 t_0} ) \sin{n \omega_0 t_0} \over b_n}\,,
\n
where $n\geq 1$. Now replacing $n$ with $m (\neq n)$, we have 
\m \label{def_C1_m}
C_1&=& 2 {\pi}^2 {a u\over L \omega_0} {{\mathcal{C}}_m \cos{m \omega_0 t_0} - {\mathcal{S}}_m \sin{m \omega_0 t_0} \over a_m}\,, \\
\label{def_C2_m}
C_2&=& 2 {\pi}^2 {a u\over L \omega_0} {({\mathcal{C}}_m + {\mathcal{S}}_m \cot{m \omega_0 t_0} ) \sin{m \omega_0 t_0} \over b_m}\,.
\n
By equaling Equation~(\ref{def_C1_n}) to Equation~(\ref{def_C1_m}) and Equation~(\ref{def_C2_n}) to Equation~(\ref{def_C2_m}), we get two independent equations for $e$ and $t_0$, which can be solved accordingly.
%

{\bf{Orbital angles $\iota$(inclination),  $\omega$(periapsis), and $\Omega$ (longitude of ascending node).}} Since we have obtained $f_0$ in the last step, to simplify the calculations, we  adapt the staring time such that $f(0)=0$, or $t_0=0$. We now define the coefficients $C_1$ and $C_2$ in Equation~(\ref{Fourier_V}) for $\Phi_{\parallel}$ and $\Phi_{\perp}$ by 
\m
C_{1 \parallel}&=& -\cos\iota \sin \omega  \sin \Omega +\cos \omega  \cos \Omega \,, \nonumber\\
C_{2 \parallel}&=&-\cos \iota \cos \omega  \sin \Omega -\sin \omega \cos \Omega \,,  \nonumber\\
C_{1 \perp}&=& \cos\iota \sin \omega  \cos \Omega +\cos \omega  \sin \Omega \,, \nonumber\\
C_{2 \perp}&=&\cos \iota \cos \omega  \cos \Omega -\sin \omega \sin \Omega \,,  
\n
and we integrate $\Phi_{\parallel}$ and $\Phi_{\perp}$ by
\begin{widetext}
\begin{subequations} 
\m \label{int_Phi_para_cos}
\int^{2\pi/\omega_0}_{0}\Phi_{\parallel} \cos{n \omega_0 t} dt&=&2 \pi u {a\over L} C_{1 \parallel} a_n {\pi\over \omega_0}=2 \pi u {a\over L} {\pi\over \omega_0}a_n(-\cos\iota \sin \omega  \sin \Omega +\cos \omega  \cos \Omega)\,, 
\n
\m  \label{int_Phi_para_sin}
\int^{2\pi/\omega_0}_{0}\Phi_{\parallel} \sin{n \omega_0 t} dt&=&2 \pi u {a\over L} C_{2 \parallel} b_n {\pi\over \omega_0}=2 \pi u {a\over L} {\pi\over \omega_0}b_n (-\cos \iota \cos \omega  \sin \Omega -\sin \omega \cos \Omega)\,, 
\n
\m  \label{int_Phi_perp_cos}
\int^{2\pi/\omega_0}_{0}\Phi_{\perp} \cos{n \omega_0 t} dt&=&2 \pi u {a\over L} C_{1 \perp} a_n {\pi\over \omega_0}=2 \pi u {a\over L} {\pi\over \omega_0}a_n(\cos\iota \sin \omega  \cos \Omega +\cos \omega  \sin \Omega)\,,
\n
\m  \label{int_Phi_perp_sin}
\int^{2\pi/\omega_0}_{0}\Phi_{\perp} \sin{n \omega_0 t} dt&=&2 \pi u {a\over L} C_{2 \perp} b_n {\pi\over \omega_0}=2 \pi u {a\over L} {\pi\over \omega_0}b_n(\cos \iota \cos \omega  \cos \Omega -\sin \omega \sin \Omega)\,,    
\n
\end{subequations}
\end{widetext}
we define 
\m
 A_n=\int^{2\pi/\omega_0}_{0}\Phi_{\perp} \sin{n \omega_0 t} dt \,, \nonumber\\
B_n=\int^{2\pi/\omega_0}_{0}\Phi_{\parallel} \cos{n \omega_0 t} dt \,, \nonumber\\
C_n=\int^{2\pi/\omega_0}_{0}\Phi_{\parallel} \sin{n \omega_0 t} dt\,, \nonumber\\
D_n=\int^{2\pi/\omega_0}_{0}\Phi_{\perp} \cos{n \omega_0 t} dt\,. 
\n

Therefore by combining Equations~(\ref{int_Phi_para_cos}) and (\ref{int_Phi_para_sin}), (\ref{int_Phi_perp_cos}) and (\ref{int_Phi_perp_sin}) we have
\begin{subequations}
\m \label{E1a}
\cos{\iota}=\cot{\Omega}\, { a_n C_n\cot{\omega + b_n B_n} \over a_n C_n -b_n B_n\cot{\omega} }\,,
\n
\m \label{E1b}
\cos{\iota}=\tan{\Omega}\, {a_n A_n \cot{\omega + b_n D_n} \over -a_n A_n +b_n D_n \cot{\omega}}\,,
\n
\end{subequations}
Similarly, by combining Equations~(\ref{int_Phi_para_cos}) and (\ref{int_Phi_perp_cos}), (\ref{int_Phi_para_sin}) and (\ref{int_Phi_perp_sin}) we have 
\begin{subequations}
\m \label{E2a}
\cos{\iota}=\cot{\omega}\, {a_n D_n -a_n B_n \tan{\Omega}\over a_n B_n + a_n D_n \tan{\Omega}}\,,
\n
\m  \label{E2b}
\cos{\iota}=- \tan{\omega}\, {b_n A_n -b_n C_n \tan{\Omega} \over b_n C_n +b_n A_n \tan{\Omega}}\,,
\n
\end{subequations}
and by combining Equations~(\ref{int_Phi_para_cos}) and (\ref{int_Phi_perp_sin}), (\ref{int_Phi_para_sin}) and (\ref{int_Phi_perp_cos}) we have
\begin{subequations}
\m \label{E3a}
\cos{\iota}={a_n A_n+b_n B_n \tan{\omega} \tan{\Omega}   \over a_n A_n \tan{\omega} \tan{\Omega} +b_n B_n}\,, 
\n 
\m  \label{E3b}
\cos{\iota}=- {a_n C_n+b_n D_n \tan{\omega} \cot{\Omega} \over a_n C_n \tan{\omega} \cot{\Omega} +b_n D_n}\,.
\n
\end{subequations}

At this point, by equaling Equation~(\ref{E1a}) and (\ref{E1b}), (\ref{E2a}) and (\ref{E2b}) we arrive at
\begin{subequations}
\m \label{eqOmega}
\tan{\Omega}=\pm \sqrt{{ (a_n C_n\cot{\omega + b_n B_n})(-a_n A_n +b_n D_n \cot{\omega}) \over(a_n C_n -b_n B_n\cot{\omega} ) (a_n A_n \cot{\omega + b_n D_n} ) }}\,,\nonumber\\
\n
\m  \label{eqomega}
\tan{\omega}=\pm \sqrt{- {(a_n D_n -a_n B_n \tan{\Omega}) (b_n C_n +b_n A_n \tan{\Omega})\over (a_n B_n + a_n D_n \tan{\Omega})(b_n A_n -b_n C_n \tan{\Omega})}}\,,
\nonumber\\
\n
\end{subequations}
 we can obtain the values of $\omega$ and $\Omega$ by solving Equations~(\ref{eqOmega}) and (\ref{eqomega}) given values of $a_n, b_n, A_n, B_n, C_n$ and $D_n$. There are constrains to these values. Firstly, they are restricted to be real, and secondly, they must satisfy the equations from (\ref{E1a}) to (\ref{E3b}), and the range of $\cos{\iota}$ is between $(-1,1)$. 

 {\bf{Ratio of $a/L$.}} Finally, we can compute $a/L$ by  integrating of $\Phi$ with either $\sin{n \omega_0 t}$ or $\cos{n \omega_0 t}$, i. e., by solving
 \m
 \int^{2\pi/\omega_0}_{0}\Phi_{\parallel} \cos{n \omega_0 t} dt&=&2 \pi u {a\over L} C_{1 \parallel} a_n {\pi\over \omega_0}\,.
 \n


\bibliographystyle{apj}
\bibliography{./reference}

\end{document}